\documentclass[pre,twocolumn,amsmath,amssymb,showkeys]{revtex4}

\usepackage{graphicx}

\newcommand{\lang}{\left\langle}
\newcommand{\rang}{\right\rangle}
\newcommand{\vecx}{\mathbf{x}}
\newcommand{\vecv}{\mathbf{v}}
\newcommand{\vhat}{\hat{\mathbf{v}}}

\begin{document}

\title{Random walk approach to the $d$-dimensional disordered Lorentz gas}

\author{Artur B. Adib}
\email{adiba@mail.nih.gov}
\affiliation{
Laboratory of Chemical Physics, NIDDK, National Institutes of Health, Bethesda, Maryland 20892-0520, USA
}

\date{\today}

\begin{abstract}
A correlated random walk approach to diffusion is applied to the disordered non-overlapping Lorentz gas. By invoking the Lu-Torquato theory for chord-length distributions in random media [J. Chem. Phys. {\bf 98}, 6472 (1993)], an analytic expression for the diffusion constant in arbitrary number of dimensions $d$ is obtained. The result corresponds to an Enskog-like correction to the Boltzmann prediction, being exact in the dilute limit, and better or nearly exact in comparison to renormalized kinetic theory predictions for all allowed densities in $d=2,3$. Extensive numerical simulations were also performed to elucidate the role of the approximations involved.
\end{abstract}


\maketitle


\section{Introduction}

The problem of computing the transport coefficients of collisional models from first-principles can be approached with a variety of methods, each with differing scope and complexity. At the most elementary level, kinetic arguments based on the concept of mean-free-path yield estimates with the correct dependence on the relevant parameters, but as one might expect miss the correct prefactors even at low densities \cite{mcquarrie-book}. Progress can be made using the Boltzmann equation and the associated Chapman-Enskog method, which give the exact low-density limit, though at the expense of substantially greater effort \cite{dorfman99}. At higher densities, renormalized kinetic theory accounts for the peculiar divergences one faces when expanding the coefficients in powers of density \cite{schram-book}, though here too the improvement comes at the cost of increased complexity in the calculation. 

For the case of the periodic Lorentz gas -- i.e. the simple model of a point particle colliding with a fixed array of spherical scatterers -- an alternate approach of great simplicity that bypasses the above kinetic route has been proposed by Machta and Zwanzig \cite{zwanzig-prl83}. These authors recognize the fact that for small enough spacings between the scatterers (i.e., in the opposite limit of the dilute approximation), the diffusing particle spends most of its time trapped inside effective ``cages'' before finding its way into an adjacent trapping region. The particle motion can thus be approximated as a random walk on a lattice, where the rate of transition is given by the cage escape rate, and consequently a simple analytic formula for the diffusion constant $D$ can be derived. The method has been subsequently extended and refined (see e.g. \cite{machta85,klages00,klages-book}), but its success remains contingent upon the prevalence of such trapping events.

\begin{figure}
\begin{center}
\includegraphics[width=220pt]{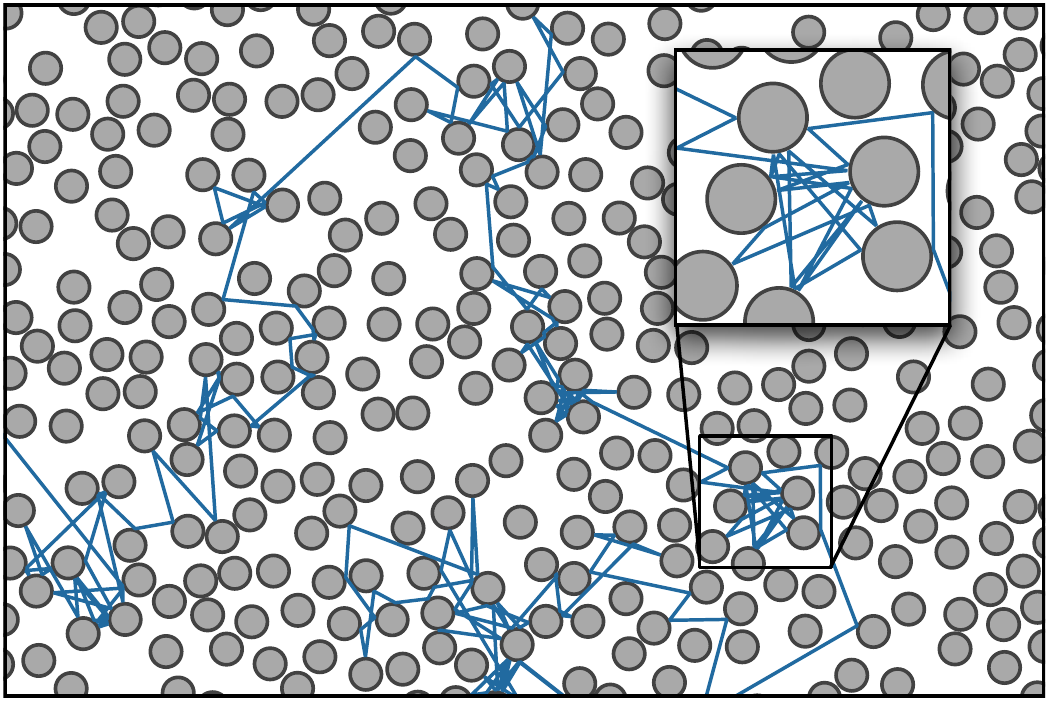} 
\caption{(Color) A typical trajectory of the non-overlapping Lorentz gas in $d=2$ at packing fraction $\phi=0.4$. The point particle propagates freely at constant velocity $\vecv$ until it suffers specular collisions with the scatterers, each of radius $R$. Inset: Representative segment of the trajectory trapped in an effective ``cage,'' an event that is ultimately responsible for correlations in the displacements $\Delta x$ between collisions. In higher number of dimensions or at lower densities, where greater voids exist between neighboring scatterers, such events are effectively suppressed.}
\label{fig:traj}
\end{center}
\end{figure}

The approach adopted in this paper---though conceptually different from that of Machta and Zwanzig---is an attempt to arrive at a result of similar simplicity and accuracy for the disordered case, without relying on the existence of kinetic traps. The method is a simple application of correlated random walks (CRW), whose main result is the ability to express $D$ very simply in terms of the correlation and moments of free-paths \cite{taylor21}. For the present problem, these can be easily computed by invoking the scattering properties of point particles by hyperspheres (Eq.~(\ref{cosine})) and the distribution of chord-lenghts in random media \cite{torquato-book}, respectively. A closely related application of CRWs has been used in the study of Knudsen diffusion in porous media \cite{derjaguin46,levitz93}. In the following, the correlated random walk method for the computation of the diffusion constant will be briefly reviewed and applied to the $d$-dimensional Lorentz gas. All calculations will be reported for the disordered non-overlapping case, where the scatterers are distributed like a liquid in equilibrium.

\section{Theory}

To begin the derivation, note that the net displacement of a particle starting from the origin consists of discrete steps $\Delta \vecx_i$ between successive collisions $i-1$ and $i$ (see Fig.~\ref{fig:traj}), so that $\vecx(n_c)=\sum_{i=1}^{n_c} \Delta \vecx_i$, where $n_c$ is the number of collisions. Thus, using the fact that $n_c$ is related to the continuous time $t$ through $t = \sum_{i=1}^{n_c} \Delta t_i = n_c \lang \Delta t \rang$ (the last equality being valid in the large $n_c$ limit) and that $\Delta t = \Delta x / v$ where $v \equiv |\vecv|$ is a constant, routine manipulations of the mean square displacement $\lang \vecx^2(n_c) \rang$ give an exact formula for $D$ expressed solely in terms of the statistics of the collision vectors, namely
\begin{equation} \label{DGK}
  D = \frac{v}{2d\lang \Delta x \rang} \left[ \lang \Delta x^2 \rang + 2 \sum_{l=1}^{\infty} \lang \Delta \vecx_0 \cdot \Delta \vecx_{l} \rang \right],
\end{equation}
where $\Delta \vecx_0$ is the displacement of the particle before its first collision. This expression can be seen as the discrete analog of the so-called Green-Kubo formula for the diffusion constant in terms of the continuous velocity correlation function (curiously, the latter was already derived by Taylor himself decades before Green and Kubo \cite{taylor21}, suggesting the more appropriate nomenclature Taylor-Green-Kubo for that result \cite{klages-book}).

The assumptions specific to the Lorentz gas that will be made (and tested) in the remainder of the paper can be cast directly in terms of the statistical properties of the collision vectors $\Delta \vecx = \Delta x \, \vhat$, where $\vhat$ is the unit velocity vector. To wit, it will be explicitly assumed that (a) their magnitudes $\Delta x$ are uncorrelated, and that (b) their orientations $\vhat$ change in a Markovian fashion. The last assumption is the simplest condition consistent with the dynamics of the gas particles, as the underlying specular collision rules manifestly introduce angular correlations between successive velocity vectors, and hence these cannot be assumed to be completely uncorrelated. As such assumptions are essentially equivalent to Boltzmann's {\em molecular chaos} hypothesis, they are satisfied for a random arrangement of scatterers at sufficiently low densities, and thus the ensuing predictions for $D$ should be exact in this limit.

\begin{figure}
\begin{center}
\includegraphics[width=220pt]{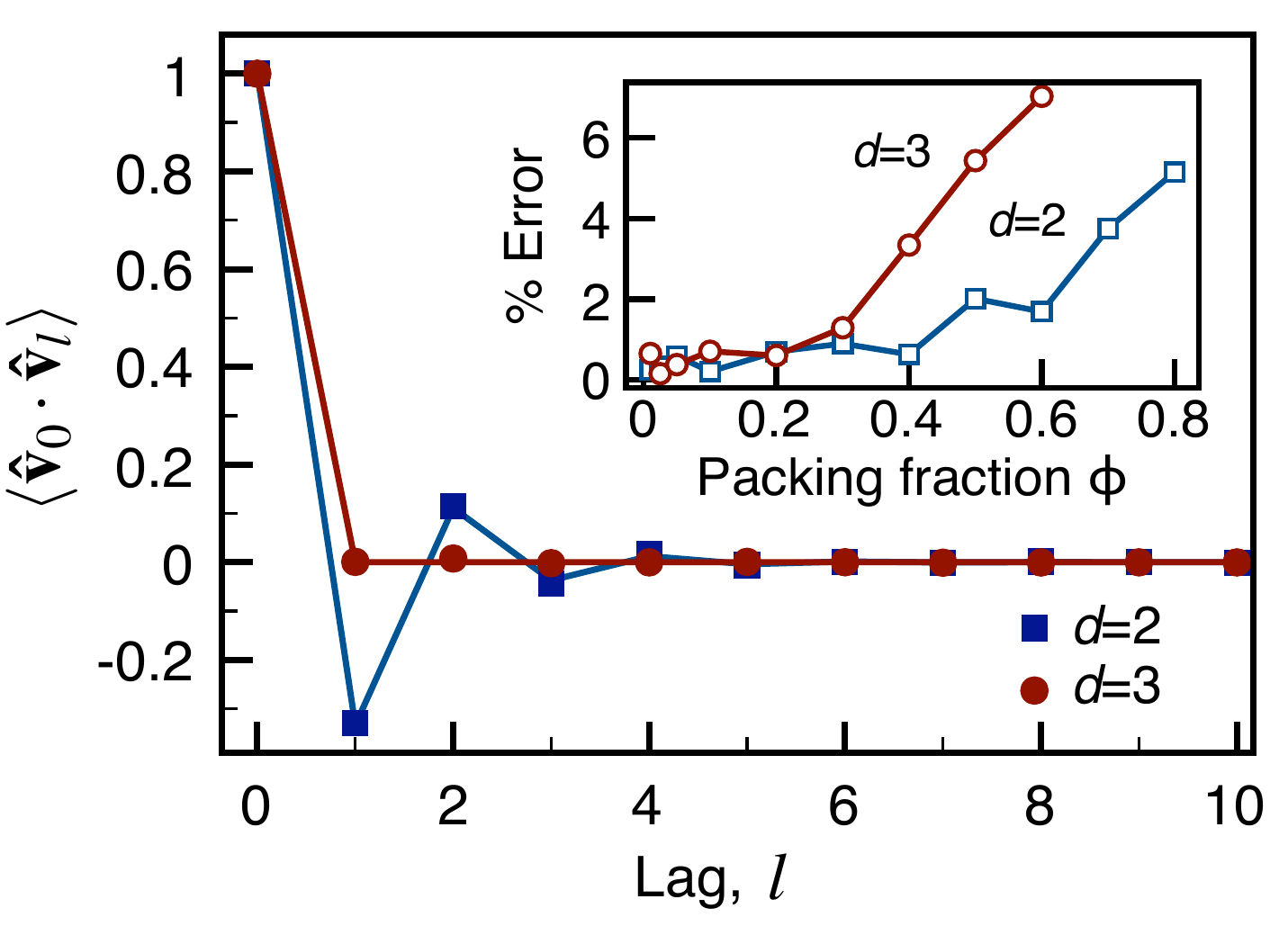} 
\caption{(Color) Discrete velocity correlation function of the Lorentz gas at packing fraction $\phi=0.1$ in $d=2$ and $d=3$, according to simulation results (main panel, symbols) and the Markovian prediction given by Eqs.~(\ref{v0dotvl}) and (\ref{cosine}) (main panel, lines). Inset: Relative error between simulation and the Markovian prediction for the integrated autocorrelation function $\sum_{l=0}^\infty \lang \vhat_0 \cdot \vhat_l \rang$.}
\label{fig:velcorr}
\end{center}
\end{figure}

With assumption (a), the infinite sum in Eq.~(\ref{DGK}) simplifies to $\lang \Delta x \rang^2 \sum_{l=1}^{\infty} \lang \vhat_0 \cdot \vhat_{l} \rang$, and in accordance with (b), we can invoke the Chapman-Kolmogorov relation to express the conditional probability $P(\vhat_l|\vhat_0)$ in terms of $P(\vhat_l|\vhat_{l-1})$ and hence obtain the recursion relation
\begin{equation}
  \lang \vhat_0 \cdot \vhat_l \rang = \lang \cos \theta \rang \, \lang \vhat_0 \cdot \vhat_{l-1} \rang,
\end{equation}
where $\lang \cos \theta \rang \equiv \lang \vhat_l\cdot\vhat_{l-1} \rang$ (this result assumes that $P(\vhat_l|\vhat_{l-1})$ has azimuthal symmetry about $\vhat_{l-1}$, as in the case of scattering by hyperspheres). This relation can be iterated to yield the simple result \footnote{As the reader might have noticed, this is a simple generalization for the case of non-fixed angles of the well known result for the correlation between bond vectors in a freely rotating ideal polymer \cite{rubinstein-book}.}
\begin{equation} \label{v0dotvl}
  \lang \vhat_0 \cdot \vhat_l \rang = \lang \cos \theta \rang^l.
\end{equation}
Now, from the specular collision rules of the gas, a direct geometric computation assuming a uniform flow of incident particles towards a $d$-dimensional spherical scatterer gives the following expression for the average scattering cosine:
\begin{equation} \label{cosine}
  \lang \cos \theta \rang = \frac{d-3}{d+1},
\end{equation}
a result that can be easily checked against the usual differential cross-sections of hard disks and hard spheres (compare with the analogous result for the case of porous media in $d=3$ \cite{derjaguin46,levitz93}). In Fig.~\ref{fig:velcorr} the above predictions for the discrete velocity correlation function (Eqs.~(\ref{v0dotvl}) and (\ref{cosine})) are compared with simulation results, illustrating in particular the robustness of the Markovian approximation even at high densities.

Use of the foregoing in Eq.~(\ref{DGK}) reduces the dependence of $D$ on the series of collision vectors to their first two marginal moments, namely
\begin{equation} \label{DRW-general}
  D = \frac{v \lang \Delta x \rang}{d} \left[ \left(\frac{\lang \Delta x^2 \rang}{2 \lang \Delta x \rang^2} - 1\right) + \frac{d+1}{4} \right].
\end{equation}
This result is a straightforward consequence of assumption (b) in correlated random walks (analogous to what was done originally in one dimension \cite{taylor21}), where the main additional ingredient is the explicit computation of the degree of correlation between adjacent $\vhat$, i.e. $\lang \vhat_l \cdot \vhat_{l-1}\rang = \lang \cos\theta \rang$, given by Eq.~(\ref{cosine}) for the present problem. Generalization to non-spherical obstacles is easily performed, provided one knows their scattering properties so that a formula similar to Eq.~(\ref{cosine}) can be obtained.

\begin{figure}
\begin{center}
\includegraphics[width=220pt]{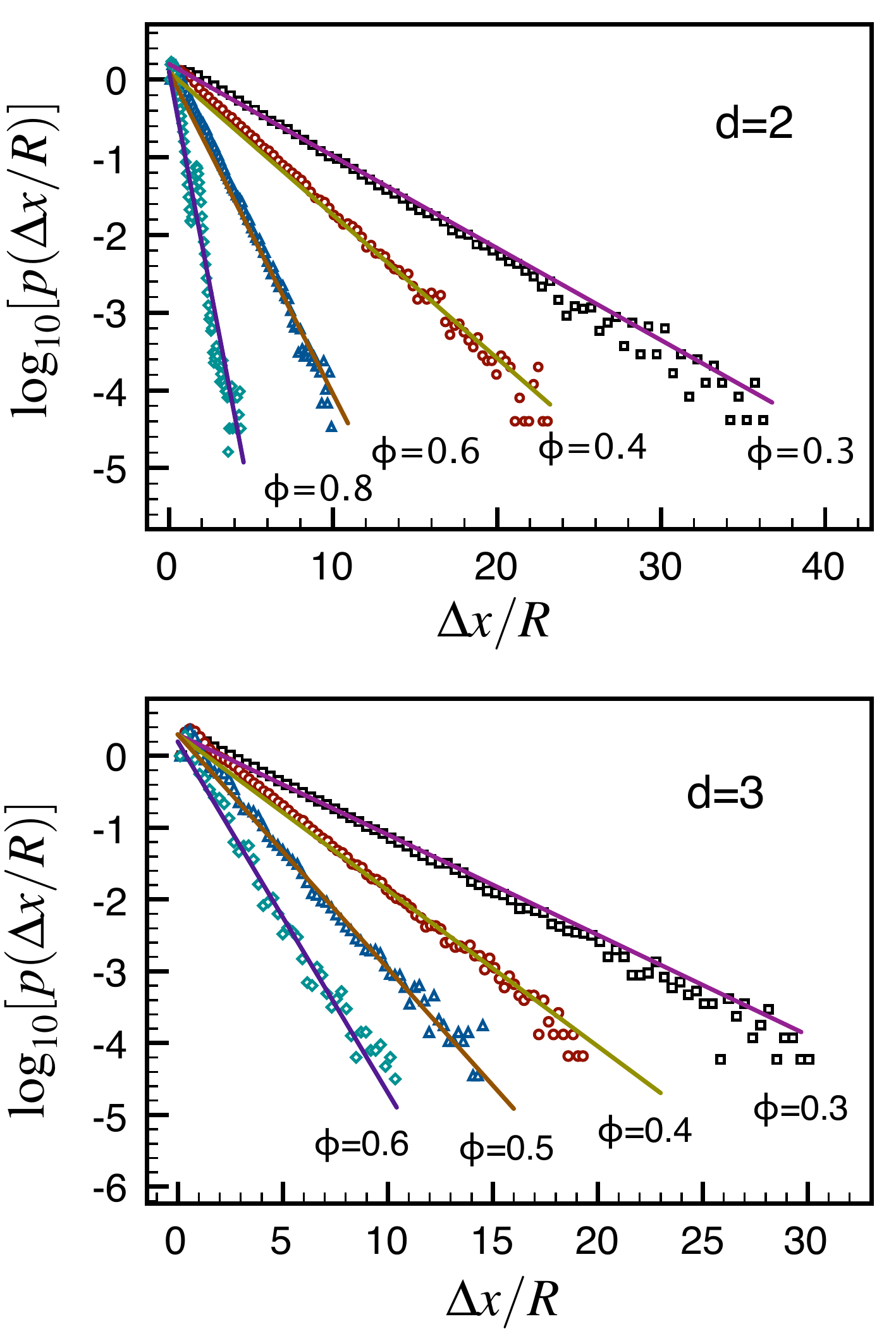} 
\caption{(Color) Distribution of collision distances $p(\Delta x/R)$ in the Lorentz gas for $d=2$ and $d=3$. The distributions are arbitrarily normalized for clarity of presentation. The symbols are simulation results at the indicated packing fractions, while the lines are exponential distributions with the exponents $\lambda$ predicted by the Lu-Torquato theory for chord-lengths (Eq.~(\ref{exponent})).}
\label{fig:deltax_both}
\end{center}
\end{figure}

\subsection{Low-density behavior}

To check the above prediction for $D$ against known results in the dilute (Boltzmann) limit, first note that in this case the $\Delta x$ are exactly exponentially distributed \cite{mcquarrie-book}, and hence the term in parenthesis in Eq.~(\ref{DRW-general}) vanishes. The first moment of this distribution is given by $1/(\sigma \rho)$, where $\sigma$ is the total cross section of the obstacles and $\rho = N/V$ is their number density. For $d=2$ and $d=3$, this yields, respectively,
\begin{equation} \label{boltzmann}
  \frac{D_B}{vR} = \frac{3\pi}{16} \frac{1}{\phi}, \quad \text{and} \quad \frac{D_B}{vR} = \frac{4}{9} \frac{1}{\phi},
\end{equation}
where $\phi$ is the packing fraction of the obstacles, and the subscript emphasizes that these results are only valid in the Boltzmann limit. These are precisely the same predictions obtained via more sophisticated methods based on, e.g., the Chapman-Enskog method or Zwanzig's operator expansion scheme \cite{weijland67,weijland68,zwanzig63}. In accordance with the general rule \cite{mcquarrie-book}, such predictions differ from those of elementary kinetic theory---namely, $D_K = v\lang \Delta x \rang/d$---by factors of order unity. (An accidental exception in the present problem is the $d=3$ case, due to the effectively randomized nature of its scattering, as can be seen by the vanishing of Eq.~(\ref{cosine})). In terms of the present framework, it can be easily seen that the source of error in elementary kinetic theory lies in its complete neglect of correlations in $\vhat$; indeed, had we made this assumption in our derivation, our expression for $D$ would have been coincident with $D_K$.

\begin{figure}
\begin{center}
\includegraphics[width=220pt]{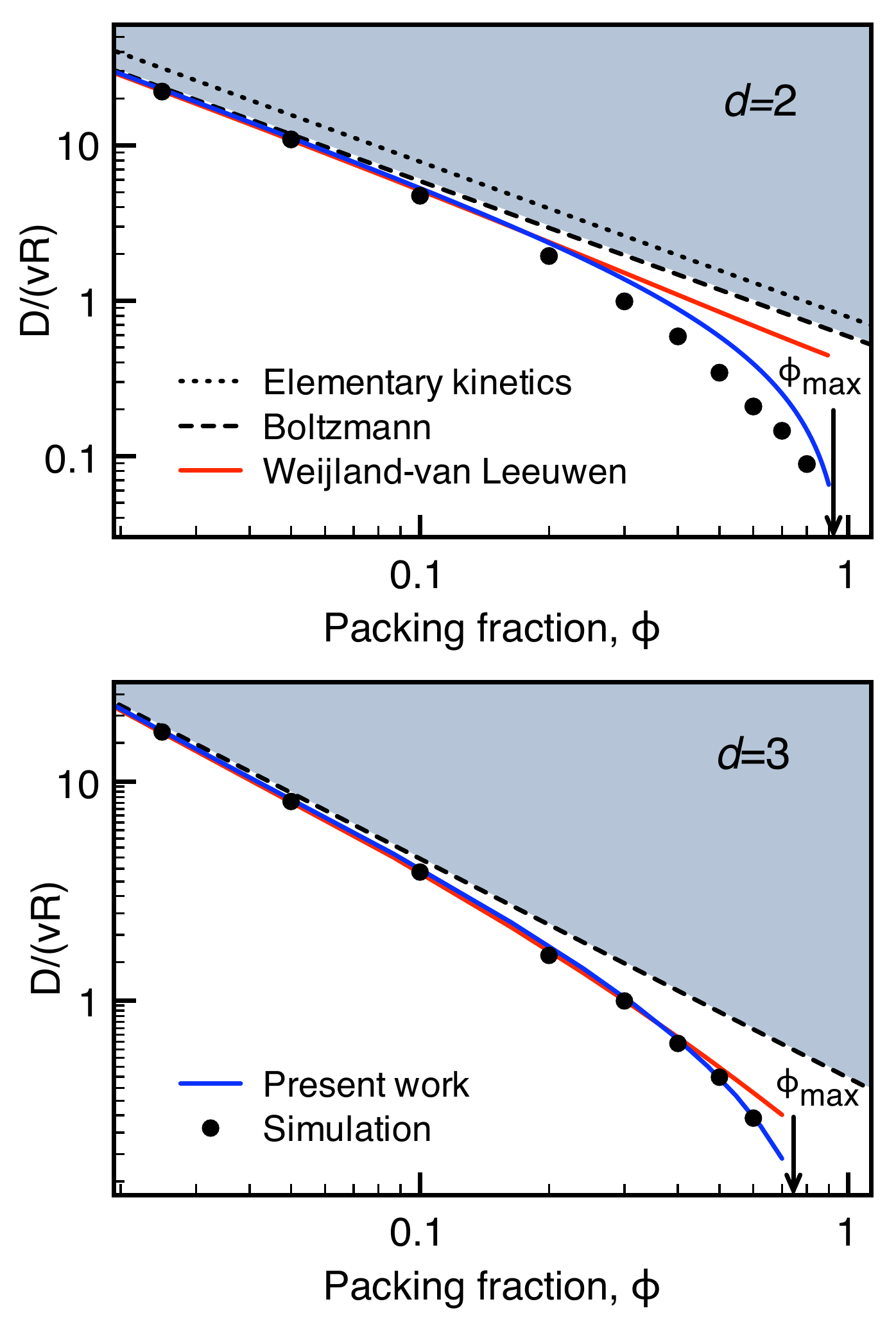} 
\caption{(Color) The diffusion constant of the Lorentz gas in $d=2$ (top) and $d=3$ (bottom) according to the various theories mentioned in the paper. Legends are the same for both figures. The elementary kinetics predictions are $D/(vR)=(\pi/4)\phi^{-1}$ for $d=2$ and $D/(vR) = (4/9)\phi^{-1}$ for $d=3$, the latter being identical to the Boltzmann prediction and hence not shown. The long-dashed lines are the Boltzmann results given by Eq.~(\ref{boltzmann}), while the red curves are from renormalized kinetic theory \cite{weijland67,weijland68}. The analytic results of this work (blue curves, closer to simulation results) are given by Eq.~(\ref{final-d23}), and are visually indistinguishable from what one obtains by using Eq.~(\ref{DRW-general}) with the required moments measured from the simulation itself (not shown). The arrows indicate the maximum packing fractions $\phi_{\text{max}} \approx 0.907$ and $\phi_{\text{max}} \approx 0.740$ for $d=2,3$, respectively \cite{torquato-book}.
}
\label{fig:D_both}
\end{center}
\end{figure}

\subsection{High-density extension}

To go beyond the Boltzmann limit, a theory for the distribution of $\Delta x$ at high densities is necessary. Here we invoke the Lu-Torquato theory for chord-length distributions in random media \cite{lutorquato93,lutorquato92}. This distribution is defined as the probability of finding an unobstructed line segment of length $\Delta x$ connecting the surfaces of any two obstacles \cite{torquato-book}. The Lu-Torquato approach is conceptually simple and elegant, and uses the scaled-particle method of liquid theory to relate chord-length distributions to the problem of finding a thin cylindrical cavity in the liquid. The central prediction for non-overlapping spheres is that the distribution is exponential, viz. $p(\Delta x) \equiv \lambda e^{-\lambda \Delta x}$, with exponent given by \cite{torquato-book}
\begin{equation} \label{exponent}
  \lambda = \frac{\omega_{d-1}}{\omega_{d}} \frac{\phi}{1-\phi} \frac{1}{R},
\end{equation}
where $\omega_d \equiv \pi^{d/2}/\Gamma(1+d/2)$ is the volume of a $d$-dimensional unit sphere. A comparison of these predictions with the present Lorentz gas simulations is shown in Fig.~\ref{fig:deltax_both}. With the exception of the salient non-exponential behavior for $\phi = 0.8$ in $d=2$ (due to excessive trapping), the theory is in excellent agreement with the numerical results.
 
\begin{figure}
\begin{center}
\includegraphics[width=220pt]{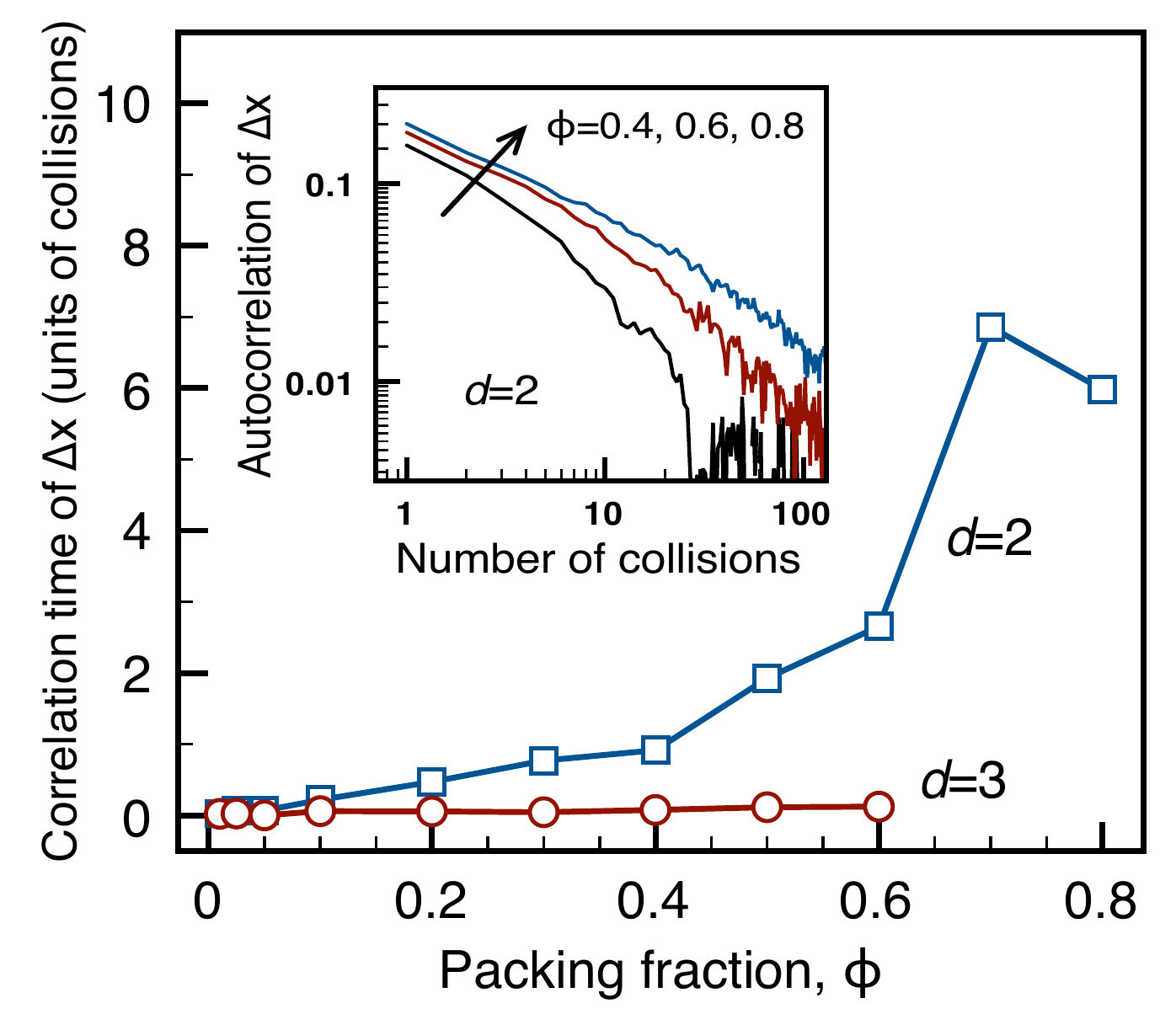} 
\caption{(Color) Correlation times of the collision distances $\Delta x$ for $d=2,3$ at various packing fractions, measured by summing the numerical results for the autocorrelation function of $\Delta x$ over all possible lags ($l=1,2,\ldots,\infty$). Inset: The autocorrelation function itself for three densities in $d=2$.}
\label{fig:autocorr}
\end{center}
\end{figure}

Finally, with Eqs.~(\ref{DRW-general}) and (\ref{exponent}) we arrive at our result for the diffusion constant of the $d$-dimensional Lorentz gas:
\begin{equation} \label{final}
  \frac{D}{vR} = \frac{d+1}{4d} \frac{\omega_d}{\omega_{d-1}} \, \frac{1-\phi}{\phi},
\end{equation}
which for the particular cases of $d=2$ and $d=3$, reads, respectively,
\begin{equation} \label{final-d23}
  \frac{D}{vR} = \frac{3\pi}{16} \, \frac{1-\phi}{\phi}, \quad \text{and} \quad \frac{D}{vR} = \frac{4}{9} \, \frac{1-\phi}{\phi}.
\end{equation}
Note that, incidentally, this result can be obtained through an {\em ad hoc} excluded-volume (i.e. Enskog-like \cite{chapman-book}) correction to the Boltzmann collision frequency, which ultimately adds an extra factor of $1-\phi$ to Eq.~(\ref{boltzmann}).

The above predictions are compared with simulation as well as with the renormalized kinetic theory of Weijland and van Leeuwen \cite{weijland67,weijland68} in Fig.~\ref{fig:D_both}. As immediately apparent, the theory is in excellent agreement with simulation for $d=3$, while the agreement for $d=2$ is less satisfactory; this can be traced back to the sensitivity of assumption (a) to the dimensionality of the problem (see Fig.~\ref{fig:autocorr}), as already anticipated in Fig.~\ref{fig:traj}.

\section{Conclusions}

To summarize, in this paper a simple analytic expression for the diffusion constant of the disordered non-overlapping Lorentz gas in arbitrary number of dimensions was derived (Eq.~(\ref{final})). This result was obtained through a simple application of correlated random walks, combining the analytic predictions of Lu and Torquato (Eq.~(\ref{exponent})) with the exact scattering properties of point particles by hard hyperspheres (Eq.~(\ref{cosine})). The simplicity and accuracy of the ensuing theory in comparison to previous renormalized kinetic methods \cite{weijland67,weijland68} are remarkable (cf. Fig.~\ref{fig:D_both}), and in this regard the goal set out in the introduction was achieved \footnote{It should be noted, however, that the present approach is not a low-density expansion, and hence it is not surprising that it does not capture the subtle logarithmic divergences of $D$ as $\phi \to 0$ predicted by the more sophisticated methods of Weijland and van Leeuwen. However, as is clear from Fig.~\ref{fig:D_both}, the theory is still in quantitative agreement for all low and intermediate densities in both $d=2,3$.}. It is hoped that these encouraging results will foster the application of similar ideas to more general models of diffusion.

\acknowledgments

The author is indebted to Attila Szabo for numerous discussions and suggestions. This research was supported by the Intramural Research Program of the NIH, NIDDK.


\begin{thebibliography}{18}
\expandafter\ifx\csname natexlab\endcsname\relax\def\natexlab#1{#1}\fi
\expandafter\ifx\csname bibnamefont\endcsname\relax
  \def\bibnamefont#1{#1}\fi
\expandafter\ifx\csname bibfnamefont\endcsname\relax
  \def\bibfnamefont#1{#1}\fi
\expandafter\ifx\csname citenamefont\endcsname\relax
  \def\citenamefont#1{#1}\fi
\expandafter\ifx\csname url\endcsname\relax
  \def\url#1{\texttt{#1}}\fi
\expandafter\ifx\csname urlprefix\endcsname\relax\def\urlprefix{URL }\fi
\providecommand{\bibinfo}[2]{#2}
\providecommand{\eprint}[2][]{\url{#2}}

\bibitem[{\citenamefont{McQuarrie}(2000)}]{mcquarrie-book}
\bibinfo{author}{\bibfnamefont{D.~A.} \bibnamefont{McQuarrie}},
  \emph{\bibinfo{title}{Statistical Mechanics}} (\bibinfo{publisher}{University
  Science}, \bibinfo{address}{Sausalito}, \bibinfo{year}{2000}).

\bibitem[{\citenamefont{Dorfman}(1999)}]{dorfman99}
\bibinfo{author}{\bibfnamefont{J.~R.} \bibnamefont{Dorfman}},
  \emph{\bibinfo{title}{An Introduction to Chaos in Nonequilibrium Statistical
  Mechanics}} (\bibinfo{publisher}{Cambridge Univ. Press},
  \bibinfo{address}{Cambridge}, \bibinfo{year}{1999}).

\bibitem[{\citenamefont{Schram}(1991)}]{schram-book}
\bibinfo{author}{\bibfnamefont{P.~P. J.~M.} \bibnamefont{Schram}},
  \emph{\bibinfo{title}{Kinetic Theory of Gases and Plasmas}}
  (\bibinfo{publisher}{Kluwer}, \bibinfo{address}{Dordrecht},
  \bibinfo{year}{1991}).

\bibitem[{\citenamefont{Machta and Zwanzig}(1983)}]{zwanzig-prl83}
\bibinfo{author}{\bibfnamefont{J.}~\bibnamefont{Machta}} \bibnamefont{and}
  \bibinfo{author}{\bibfnamefont{R.}~\bibnamefont{Zwanzig}},
  \bibinfo{journal}{Phys. Rev. Lett.} \textbf{\bibinfo{volume}{50}},
  \bibinfo{pages}{1959} (\bibinfo{year}{1983}).

\bibitem[{\citenamefont{Machta and Moore}(1985)}]{machta85}
\bibinfo{author}{\bibfnamefont{J.}~\bibnamefont{Machta}} \bibnamefont{and}
  \bibinfo{author}{\bibfnamefont{S.~M.} \bibnamefont{Moore}},
  \bibinfo{journal}{Phys. Rev. A} \textbf{\bibinfo{volume}{32}},
  \bibinfo{pages}{3164} (\bibinfo{year}{1985}).

\bibitem[{\citenamefont{Klages and Dellago}(2000)}]{klages00}
\bibinfo{author}{\bibfnamefont{R.}~\bibnamefont{Klages}} \bibnamefont{and}
  \bibinfo{author}{\bibfnamefont{C.}~\bibnamefont{Dellago}},
  \bibinfo{journal}{J. Stat. Phys.} \textbf{\bibinfo{volume}{101}},
  \bibinfo{pages}{145} (\bibinfo{year}{2000}).

\bibitem[{\citenamefont{Klages}(2007)}]{klages-book}
\bibinfo{author}{\bibfnamefont{R.}~\bibnamefont{Klages}},
  \emph{\bibinfo{title}{Microscopic Chaos, Fractals and Transport in
  Nonequilibrium Statistical Mechanics}} (\bibinfo{publisher}{World
  Scientific}, \bibinfo{address}{Hackensack}, \bibinfo{year}{2007}).

\bibitem[{\citenamefont{Taylor}(1921)}]{taylor21}
\bibinfo{author}{\bibfnamefont{G.~I.} \bibnamefont{Taylor}},
  \bibinfo{journal}{Proc. London Math. Soc.} \textbf{\bibinfo{volume}{20}},
  \bibinfo{pages}{196} (\bibinfo{year}{1921}).

\bibitem[{\citenamefont{Torquato}(2002)}]{torquato-book}
\bibinfo{author}{\bibfnamefont{S.}~\bibnamefont{Torquato}},
  \emph{\bibinfo{title}{Random Heterogeneous Materials: Microstructure and
  Macroscopic Properties}} (\bibinfo{publisher}{Springer},
  \bibinfo{address}{New York}, \bibinfo{year}{2002}).

\bibitem[{\citenamefont{Derjaguin}(1946)}]{derjaguin46}
\bibinfo{author}{\bibfnamefont{B.}~\bibnamefont{Derjaguin}},
  \bibinfo{journal}{C. R. Acad. Sci. URSS} \textbf{\bibinfo{volume}{53}},
  \bibinfo{pages}{623} (\bibinfo{year}{1946}).

\bibitem[{\citenamefont{Levitz}(1993)}]{levitz93}
\bibinfo{author}{\bibfnamefont{P.}~\bibnamefont{Levitz}}, \bibinfo{journal}{J.
  Phys. Chem.} \textbf{\bibinfo{volume}{97}}, \bibinfo{pages}{3813}
  (\bibinfo{year}{1993}).

\bibitem[{\citenamefont{{Van Leeuwen} and Weijland}(1967)}]{weijland67}
\bibinfo{author}{\bibfnamefont{J.~M.~J.} \bibnamefont{{Van Leeuwen}}}
  \bibnamefont{and} \bibinfo{author}{\bibfnamefont{A.}~\bibnamefont{Weijland}},
  \bibinfo{journal}{Physica} \textbf{\bibinfo{volume}{36}},
  \bibinfo{pages}{457} (\bibinfo{year}{1967}).

\bibitem[{\citenamefont{Weijland and {Van Leeuwen}}(1968)}]{weijland68}
\bibinfo{author}{\bibfnamefont{A.}~\bibnamefont{Weijland}} \bibnamefont{and}
  \bibinfo{author}{\bibfnamefont{J.~M.~J.} \bibnamefont{{Van Leeuwen}}},
  \bibinfo{journal}{Physica} \textbf{\bibinfo{volume}{38}}, \bibinfo{pages}{35}
  (\bibinfo{year}{1968}).

\bibitem[{\citenamefont{Zwanzig}(1963)}]{zwanzig63}
\bibinfo{author}{\bibfnamefont{R.}~\bibnamefont{Zwanzig}},
  \bibinfo{journal}{Phys. Rev.} \textbf{\bibinfo{volume}{129}},
  \bibinfo{pages}{486} (\bibinfo{year}{1963}).

\bibitem[{\citenamefont{Lu and Torquato}(1993)}]{lutorquato93}
\bibinfo{author}{\bibfnamefont{B.}~\bibnamefont{Lu}} \bibnamefont{and}
  \bibinfo{author}{\bibfnamefont{S.}~\bibnamefont{Torquato}},
  \bibinfo{journal}{J. Chem. Phys.} \textbf{\bibinfo{volume}{98}},
  \bibinfo{pages}{6472} (\bibinfo{year}{1993}).

\bibitem[{\citenamefont{Lu and Torquato}(1992)}]{lutorquato92}
\bibinfo{author}{\bibfnamefont{B.}~\bibnamefont{Lu}} \bibnamefont{and}
  \bibinfo{author}{\bibfnamefont{S.}~\bibnamefont{Torquato}},
  \bibinfo{journal}{Phys. Rev. A} \textbf{\bibinfo{volume}{45}},
  \bibinfo{pages}{922} (\bibinfo{year}{1992}).

\bibitem[{\citenamefont{Chapman and Cowling}(1970)}]{chapman-book}
\bibinfo{author}{\bibfnamefont{S.}~\bibnamefont{Chapman}} \bibnamefont{and}
  \bibinfo{author}{\bibfnamefont{T.~G.} \bibnamefont{Cowling}},
  \emph{\bibinfo{title}{The Mathematical Theory of Non-Uniform Gases}}
  (\bibinfo{publisher}{Cambridge}, \bibinfo{address}{Cambridge},
  \bibinfo{year}{1970}), \bibinfo{edition}{3rd} ed.

\bibitem[{\citenamefont{Rubinstein and Colby}(2003)}]{rubinstein-book}
\bibinfo{author}{\bibfnamefont{M.}~\bibnamefont{Rubinstein}} \bibnamefont{and}
  \bibinfo{author}{\bibfnamefont{R.~H.} \bibnamefont{Colby}},
  \emph{\bibinfo{title}{Polymer Physics}} (\bibinfo{publisher}{Oxford},
  \bibinfo{address}{Oxford}, \bibinfo{year}{2003}).

\end{thebibliography}
\end{document}